\documentclass[conference]{IEEEtran}
\IEEEoverridecommandlockouts

\usepackage{cite}
\usepackage{amsmath,amsthm,amssymb,amsfonts}
\usepackage{algorithmic}
\usepackage{graphicx}
\usepackage{textcomp}
\usepackage{xcolor}
\def\BibTeX{{\rm B\kern-.05em{\sc i\kern-.025em b}\kern-.08em
		T\kern-.1667em\lower.7ex\hbox{E}\kern-.125emX}}

\newcommand{\C}{{\mathbb{C}}}

\newcommand{\N}{\mathbb{N}}
\newcommand{\Z}{\mathbb{Z}}

\newtheorem{theorem}{Theorem}
\newtheorem{lemma}[theorem]{Lemma}
\newtheorem{proposition}[theorem]{Proposition}
\newtheorem{definition}[theorem]{Definition}

\newtheorem{remark}[theorem]{Remark}

\usepackage{tikz}
\usepackage{textcomp}
\usepackage{hyperref}
\usepackage{lipsum}

\newcommand\copyrighttext{%
	\footnotesize \textcopyright 2025 IEEE.  Personal use of this material is permitted.  Permission from IEEE must be obtained for all other uses, in any current or future media, including reprinting/republishing this material for advertising or promotional purposes, creating new collective works, for resale or redistribution to servers or lists, or reuse of any copyrighted component of this work in other works.
	}

\newcommand\copyrightnotice{%
	\begin{tikzpicture}[remember picture,overlay]
		\node[anchor=north,yshift=-10pt] at (current page.north) {\fbox{\parbox{\dimexpr\textwidth-\fboxsep-\fboxrule\relax}{\copyrighttext}}};
	\end{tikzpicture}%
}

\newcommand\citationtext{%
	\footnotesize H. Führ and T. Gilles, ``The Restricted Isometry Property for Measurements from Group Orbits'', 2025 International Conference on Sampling Theory and Applications (SampTA), 2025, pp. 1-5, 
	DOI: \href{https://doi.org/10.1109/SampTA64769.2025.11133511}{10.1109/SampTA64769.2025.11133511}.
}

\newcommand\citationnotice{%
	\begin{tikzpicture}[remember picture,overlay]
		\node[anchor=south,yshift=10pt] at (current page.south) {\fbox{\parbox{\dimexpr\textwidth-\fboxsep-\fboxrule\relax}{\citationtext}}};
	\end{tikzpicture}%
}

\begin{document}
	
	\title{The Restricted Isometry Property for Measurements from Group Orbits\\
		\thanks{Funded by the Deutsche Forschungsgemeinschaft (DFG, German Research Foundation) – SFB 1481 – 442047500}
	}
	
	\author{\IEEEauthorblockN{Hartmut Führ}
		\IEEEauthorblockA{\textit{Chair of Geometry and Analysis} \\
			\textit{RWTH Aachen University}\\
			Aachen, Germany \\
			fuehr@mathga.rwth-aachen.de}
		\and
		\IEEEauthorblockN{Timm Gilles}
		\IEEEauthorblockA{\textit{Chair of Geometry and Analysis} \\
			\textit{RWTH Aachen University}\\
			Aachen, Germany \\
			gilles@mathga.rwth-aachen.de}
	}
	
	\maketitle
	\citationnotice
	\copyrightnotice
	
	\pagestyle{plain}
	
	\begin{abstract}
		It is known that sparse recovery by measurements from random circulant matrices provides good recovery bounds. We generalize this to measurements that arise as a random orbit of a group representation for some finite group G. We derive estimates for the number of measurements required to guarantee the restricted isometry property with high probability. Following this, we present several examples highlighting the role of appropriate representation-theoretic assumptions.
	\end{abstract}

\section{Introduction}
We are interested in recovering an $s$-sparse signal $x\in\C^n$, i.e. $\Vert x\Vert_0:=\vert\{ j\in\{1,\ldots, n\}\mid x_j\neq 0\}\vert\leq s$, which is measured by the linear measurement process
\begin{align*}
	y=\Phi x
\end{align*}
from as few measurements as possible. Here, $\Phi\in\C^{m\times n}$ is called \textit{measurement matrix} and $y\in\C^m$ is the vector of observed measurements. For a matrix $\Phi\in\C^{m \times n}$ and $1\leq s\leq n$, the restricted isometry constant $\delta_s$ is the smallest $\delta\geq 0$ such that
\[
(1-\delta)\Vert x\Vert_2^2\leq \Vert \Phi x\Vert_2^2 \leq(1+\delta) \Vert x\Vert_2^2 
\]
for all $s$-sparse vectors $x\in\C^n$. It is well known that if $\Phi$ satisfies the above condition, also called restricted isometry property, for a small enough constant $\delta_s$, there exist efficient algorithms that recover any $s$-sparse signal $x$, see \cite{candes2005decoding}, \cite{mo2011new}, \cite{foucart2011hard}. It has been shown in \cite{candes2006near} that if $\Phi$ is a Gaussian random matrix, i.e its entries are independent and follow a normal distribution with mean 0 and variance 1, and
\begin{align}\label{eq:optimal_bound_measurements}
	m\geq C  \delta^{-2} s \ln\left(\frac{n}{s}\right)
\end{align} 
holds, then $\delta_s\leq \delta$ with high probability. Moreover, all so far known measurement matrices that provide optimal bounds, i.e. $m$ scales as in (\ref{eq:optimal_bound_measurements}) \cite{foucart2013mathematical}, are random. However, in practice, one usually has more structured measurement processes. Specific choices of structured measurement matrices have been studied, such as Random Partial Fourier Matrices in \cite{candes2006robust}, or Partial Random Circulant Matrices and Time-Frequency Structured Random Matrices in \cite{krahmer2014suprema}.\par
In this work however, we aim for a more general setup of structured measurement matrices that arise from inner products of the signal $x$ with elements of some group orbit. To precisely state the measurement matrix $\Phi$ of interest, we first need some notation: Let $G$ be a finite group. We define a projective representation $\pi\colon G\to \text{GL}(\C^n)$ to be a mapping such that for all $g,h\in G$ there exists $\lambda(g,h)\in\C$ with $\vert \lambda(g,h)\vert=1$ and 
\[
\pi(g)\pi(h)=\lambda(g,h)\pi(gh).
\]
$\pi$ is called unitary if $\pi(g)$ is unitary for every $g\in G$. Lastly, $\pi$ is a representation if $\lambda(g,h)=1$ for all $g,h\in G$. For a more detailed introduction to group representations see \cite{terras1999fourier}. Lastly, we call a random vector $X$ on $\C^n$ $L$-subgaussian if $\mathbb{E}X=0$, $\mathbb{E}\vert \langle X,\theta\rangle\vert^2=1$ for every $\theta\in S^{n-1}$ and
\begin{align*}
	\mathbb{P}(\vert\langle X,\theta\rangle\vert\geq t)\leq 2\exp\left(-\frac{t^2}{2L^2}\right)
\end{align*} 
for every $\theta\in S^{n-1}$ and any $t>0$. With this notation at hand we are ready to define the measurement matrix $\Phi$.

\begin{definition}\label{def:measurement_mat}
	Let $\pi\colon G\to \text{GL}(\C^n)$ be a unitary (projective) representation. Fix some sampling set $\Omega=\{\omega_1,\ldots,\omega_m\}\subset G$ of size $\vert \Omega\vert=m$. Let $\xi$ be a random vector with independent mean 0, variance 1, and $L$-subgaussian entries. Set
	\begin{align*}
		\Phi=\frac{1}{\sqrt{m}}\, R_{\Omega}\big(\pi(g) \xi\big) ^{\ast}_{g\in G}\in \C^{m\times n}
	\end{align*}
	where $(\pi(g)\xi)_{g\in G}$ denotes the matrix in $\C^{n\times n}$ which columns are the vectors $\pi(g)\xi$ and $R_{\Omega}\colon \C^G\to\C^m$ restricts a vector to its entries in $\Omega$, i.e. $(R_{\Omega}y)_l=y(\omega_l)$.
\end{definition}

A special case of Definition \ref{def:measurement_mat} are the Partial Random Circulant Matrices mentioned before. Therefore, our main result, Theorem \ref{mainresult}, generalizes \cite[Theorem 4.1]{krahmer2014suprema}. We will make this precise in Section \ref{sect:constant}.

\section{The restricted isometry property}\label{sect:rip}

Intuitively, different representations should give different sparse recovery results. For example, it can be shown that sparse recovery is not possible when considering the trivial representation $\pi(g)=\text{Id}$. To measure this performance difference between representations, we have added the restriction (\ref{main_annahme}) in Theorem \ref{mainresult}. Here, the size of the constant $C_{\Omega,\pi}$ plays a crucial role. It essentially characterizes how well a given representation performs in sparse recovery. In Section \ref{sect:constant}, we present examples for the choice of $C_{\Omega,\pi}$.\par
The proof idea of Theorem \ref{mainresult} was first seen in \cite{krahmer2014suprema}.

\begin{theorem}\label{mainresult}
	Let $\Phi$ be as in Definition \ref{def:measurement_mat}. Further, let $C_{\Omega,\pi}>0$ be a constant, which can depend on all of $\Omega$'s characteristics except its size $m$, such that
	\begin{align}\label{main_annahme}
		\sup_{1\leq j\leq n, y\in\C^n} \Big\Vert \frac{1}{\sqrt{m}}\, R_{\Omega}\big(\pi(g) y\big) ^{\ast}_{g\in G} e_j 	\Big\Vert_{2} \leq \sqrt{\frac{C_{\Omega,\pi}}{m}} \Vert y\Vert_2
	\end{align}
	holds. If, for $s\leq n$ and $\delta,\eta \in(0,1)$,
	\begin{align}\label{gl:absm}
		m\geq c \delta^{-2}s \,C_{\Omega,\pi} \max\big\{\!(\log(s\, C_{\Omega,\pi})\log n)^2,\,\log(\eta^{-1})\!\big\},
	\end{align}
	then with probability at least $1-2\eta$, the restricted isometry constant of $\Phi$ satisfies $\delta_s\leq \delta$. Here, $c>0$ only depends on $L$.
\end{theorem}
\begin{IEEEproof}
	We start by explaining the main idea of the proof: The restricted isometry constant of $\Phi$ can be rewritten as supremum of a chaos process for some set of matrices $\mathcal{A}$. For this type of process, we can then use the concentration inequality of the following theorem.
	
	\begin{theorem}\label{th:concentration_inequ}\cite[Theorem 3.1]{krahmer2014suprema}
		Let $\mathcal{A}$ be a set matrices, and let $\xi$ be a random vector whose entries $\xi_j$ are independent, mean 0, variance 1 and $L$-subgaussian random variables. Define $d_{2\to 2}(\mathcal{A})=\sup_{A\in\mathcal{A}}\Vert A\Vert_{2\to 2}$ and $d_{F}(\mathcal{A})=\sup_{A\in\mathcal{A}}\Vert A\Vert_{F}$. By $\gamma_2(\mathcal{A},\Vert\cdot\Vert_{2\to2})$ we denote Talagrand's functional (see \cite[Definition 2.1]{krahmer2014suprema}). Set
		\begin{align*}
			E&=\gamma_2(\mathcal{A},\Vert\cdot\Vert_{2\to 2})(\gamma_2(\mathcal{A},\Vert\cdot\Vert_{2\to 2})+d_F(\mathcal{A}))\\
			&\,\,\,\,\,\,+d_F(\mathcal{A})d_{2\to 2}(\mathcal{A}),\\
			V&=d_{2\to 2}(\mathcal{A})(\gamma_2(\mathcal{A},\Vert\cdot\Vert_{2\to 2})+d_F(\mathcal{A})),\\
			U&=d^2_{2\to 2}(\mathcal{A}).
		\end{align*}
		Then, for $t>0$,
		\begin{align*}
			&\mathbb{P}\left(\sup_{A\in\mathcal{A}}\left\vert\Vert A\xi\Vert_2^2-\mathbb{E}\Vert A\xi\Vert_2^2\right\vert \geq c_1E+t \right)\\
			&\leq 2\exp\left(-c_2\min\left\{\frac{t^2}{V^2},\frac{t}{U}\right\}\right).
		\end{align*}
		The constants $c_1,c_2>0$ depend only on $L$.
	\end{theorem}
	
	Then, it will be sufficient to find suitable bounds for $d_{2\to 2}(\mathcal{A})$, $d_{F}(\mathcal{A})$ and $\gamma_{2}(\mathcal{A},\Vert\cdot\Vert_{2\to 2})$. Let us start with some notation. Let
	\[
	D_{s,n}=\{x\in\C^n \mid \Vert x\Vert_2\leq 1, \Vert x\Vert_0\leq s\}
	\]
	and for every $x\in\C^n$ define the matrix 
	\begin{align*}
		A_x=\frac{1}{\sqrt{m}} R_{\Omega}\left(\pi(g)^{\ast}x\right)^{\ast}_{g\in G}\in \C^{m \times n}.
	\end{align*}
	Further, set
	$\mathcal{A}=\{ A_x \mid x\in D_{s,n}\}$. Since the entries of $\xi$ are independent and have mean 0 and variance 1, it holds
	\begin{align*}
		\mathbb{E} \vert \langle \xi, y\rangle \vert^2=\mathbb{E}\sum_{i,j=1}^n y_i \overline{y_j} \overline{ \xi_i} \xi_j= \sum_{i,j=1}^n y_i \overline{y_j} \underbrace{\mathbb{E} \overline{ \xi_i} \xi_j}_{=\delta_{ij}}=\Vert y\Vert_2^2
	\end{align*}
	for all $y\in\C^n$. Together with the fact that the representation $\pi$ is unitary we get
	\begin{align*}
		\mathbb{E}\Vert A_x\xi \Vert_2^2&=\mathbb{E}\sum_{g \in \Omega} \frac{1}{m} \vert \langle \xi, \pi(g)^{\ast} x\rangle\vert^2=\frac{1}{m}\sum_{g \in \Omega} \mathbb{E}\vert \langle \xi,\pi(g)^{\ast} x\rangle\vert^2\\
		&=\frac{1}{m}\sum_{g \in \Omega} \Vert \pi(g)^{\ast} x\Vert^2_2 = \Vert x\Vert^2_2
	\end{align*}
	for all $x\in\C^n$. Now we are able to rewrite the restricted isometry constant as supremum of a chaos process, i.e
	\begin{align*}
		\delta_s&=\sup_{x\in D_{s,n}} \Big| \Vert \Phi x \Vert_2^2-\Vert x\Vert^2_2 \Big|=\sup_{x\in D_{s,n}} \Big| \Vert A_x \xi\Vert_2^2-\mathbb{E} \Vert A_x \xi\Vert^2_2\Big|.
	\end{align*}
	For better readability, we prove bounds for the above quantities separately in the following lemmata.
	
	\begin{lemma}\label{lem:estimate_A_x_2to2}
		It holds
		\begin{align*}
			\Vert A_x\Vert_{2\to2}\leq \sqrt{\frac{s\, C_{\Omega,\pi}}{m}}\Vert x\Vert_2
		\end{align*}
		for all $x\in\C^n$ with $\Vert x\Vert_0\leq s$. Thus, $d_{2\to 2}(\mathcal{A})\leq \sqrt{\frac{s\, C_{\Omega,\pi}}{m}}$.
	\end{lemma}
	\begin{IEEEproof}
		Let $x\in\C^n$ with $\Vert x\Vert_0\leq s$. We can rewrite the $s$-sparse vector $x$ as $\sum_{j\in J} \alpha_j e_j$ with $\vert J\vert \leq s$. Then, the triangle inequality and Cauchy-Schwarz inequality give
		\begin{align*}
			\Big\Vert \frac{1}{\sqrt{m}} R_{\Omega}(\pi(g)y)^{\!\ast}_{\!g\in G} x\Big\Vert_2&\leq \sum_{j \in J} \vert \alpha_j\vert\Big\Vert \frac{1}{\sqrt{m}} R_{\Omega}(\pi(g)y)^{\!\ast}_{\!g\in G}  e_j\Big\Vert_2\\
			& \leq \sum_{j\in J} \vert \alpha_j\vert \sqrt{\frac{C_{\Omega,\pi}}{m}}\Vert y\Vert_2 \\
			& \leq \sqrt{\frac{C_{\Omega,\pi}}{m}} \Vert y\Vert_2 \sqrt{\vert J\vert} \Big( \sum_{j\in J} \vert \alpha_j\vert^2\Big)^{\frac{1}{2}}\\
			&\leq \sqrt{\frac{s\, C_{\Omega,\pi}}{m}} \Vert y\Vert_2 \Vert x\Vert_2
		\end{align*} 
		for all $y\in \C^n$ and $g\in\Omega$. Since $\vert\langle y,\pi(g)^{\ast}x\rangle \vert=\vert \langle x,\pi(g)y\rangle \vert$ holds for all $y\in\C^n$ and $g\in\Omega$, it follows that
		\begin{align*}
			\Vert A_x\Vert_{2\to2} &=\!\sup_{\Vert y\vert_2=1}\!\Big\Vert \frac{1}{\sqrt{m}}\, R_{\Omega}\big(\pi(g) y\big) ^{\ast}_{g\in G} x 	\Big\Vert_{2}\leq \sqrt{\!\frac{s\, C_{\Omega,\pi}}{m}\!} \Vert x\Vert_2. 
		\end{align*}
		Hence, $d_{2\to 2}(\mathcal{A})\leq \sqrt{\frac{s\, C_{\Omega,\pi}}{m}}$.
	\end{IEEEproof}

	\begin{lemma}\label{lem_estimate_A_x_F}
		It holds $d_F(\mathcal{A})=1$.
	\end{lemma}	
	\begin{IEEEproof}
		Let $x\in\C^n$ with $\Vert x\Vert_0\leq s$. Then, $\pi$ being unitary implies
		\begin{align*}
			\Vert A_x \Vert_F^2&=\frac{1}{m} \Big\Vert R_{\Omega} \left(\pi(g)^{\ast}x\right)^{\ast}_{g\in G}\Big\Vert_F^2=\frac{1}{m} \Big\Vert \left(\pi(g)^{\ast}x\right)_{g\in \Omega}\Big\Vert_F^2\\
			&=\frac{1}{m} \Big\Vert \big(\pi(g_1)^{\ast} x\mid \cdots\mid \pi(g_m)^{\ast}x\big) \Big\Vert_F^2\\
			&=\frac{1}{m} \sum_{i=1}^m \Vert \pi(g_i)^{\ast} x\Vert_2^2 =\frac{1}{m} \sum_{i=1}^m \Vert x\Vert_2^2=\Vert x\Vert_2^2.
		\end{align*}
		Thus, $d_F(\mathcal{A})=1$.
	\end{IEEEproof}
	
	Lastly, we have to bound the term $\gamma_{2}(\mathcal{A},\Vert\cdot\Vert_{2\to 2})$. In \cite{vershynin2018high}, a bound in form of a Dudley integral (first considered in \cite{dudley1967sizes})
	\begin{align*}
		\gamma_{2}(\mathcal{A},\Vert\cdot\Vert_{2\to 2})\lesssim \int\limits_{0}^{d_{2\to 2}(\mathcal{A})}\sqrt{\log \mathcal{N}(\mathcal{A},\Vert\cdot\Vert_{2\to 2},t)}\,\text{d}t
	\end{align*}
	was shown. Here, $A\lesssim B$ means that there exists a universal constant $c_1>0$ such that $A\leq c_1 B$. Usually, one then proves two bounds for the covering number $\mathcal{N}(\mathcal{A},\Vert\cdot\Vert_{2\to 2},t)$, which is the minimal number of open balls in $(\mathcal{A},\Vert\cdot\Vert_{2\to 2}$) of radius $t$ that is needed to cover $\mathcal{A}$. The first bound uses the empirical method of Maurey \cite{carl1985inequalities} and the second one a volumetric argument. This approach has also been used in several similar settings, see \cite{rudelson2008sparse} or \cite{rauhut2010compressive}. 
	
	\begin{lemma}\label{lem:covnumbabs}
		Let $t>0$. We have the following inequalities:
		\begin{align*}
			\log \mathcal{N}(\mathcal{A},\Vert\cdot\Vert_{2\to2},t)\lesssim \frac{{s\, C_{\Omega,\pi}}}{m} \frac{1}{t^2}{\log (n)} \log(4n)
		\end{align*}
		and
		\begin{align*}
			\log\mathcal{N}(\mathcal{A},\Vert\cdot\Vert_{2\to2},t)\lesssim s\bigg(&\log  \Big(1+2\frac{\sqrt{s\, C_{\Omega,\pi}}}{\sqrt{m}\,t}\Big)\\
			& +\log\left(\frac{en}{s}\right)\bigg).
		\end{align*}
	\end{lemma}
	\begin{IEEEproof}
		Let $t>0$. Define a set of matrices by
		\begin{align*}
			B=\{A_{\pm\sqrt{2}e_1},\ldots,A_{\pm\sqrt{2}e_n},A_{\pm\sqrt{2}i e_1},\ldots,A_{\pm\sqrt{2}i e_n} \}.
		\end{align*}
		It is easy to check that
		\begin{align*}
			D_{s,n}\subseteq\! \sqrt{s}\,\text{conv}(\pm\sqrt{2}e_1,\ldots,\pm\sqrt{2}e_n,\pm\sqrt{2}ie_1,\ldots,\pm\sqrt{2}ie_n)
		\end{align*}
		holds, where $\text{conv}()$ denotes the convex hull. Combining this with the fact that
		\begin{align*}
			A_{\sum_{j=1}^N\alpha_j z^j}=\sum_{j=1}^N \overline{\alpha_j}\, A_{z^j}
		\end{align*}
		for all $\alpha_1,\ldots,\alpha_N\in\C$ and $z^1,\ldots,z^N\in\C^n$, gives the relation $\mathcal{A}\subset \sqrt{s}\,\text{conv}(B)$. Thus,
		\begin{align}\label{eq:cov_numb_conv_subset}
			\mathcal{N}(\mathcal{A},\Vert\cdot\Vert_{2\to2},t)\leq \mathcal{N}\left(\text{conv}(B),\Vert\cdot\Vert_{2\to2},\frac{1}{\sqrt{s}}t\right).
		\end{align}
		To bound the right side we use \cite[Lemma 4.2]{krahmer2014suprema} which is based on the empirical method of Maurey. Let $N\in\N$, $(A_1,\ldots,A_N)\in B^L$ and  let $(e_j)_{j=1}^N$ be a random vector with independent Rademacher distributed entries. It holds
		\begin{align*}
			\mathbb{E}_{\epsilon}\Big\Vert \sum_{j=1}^N \epsilon_j A_j\Big\Vert_{2\to 2}
			&\,\,{\lesssim} \sqrt{\log(n)} \max\Big\{\Big\Vert \sum_{j=1}^N A_{j}^{\ast} A_{j}\Big\Vert_{2\to 2},\\
			&\,\,\,\,\,\,\,\,\,\,\,\,\,\,\,\,\,\,\,\,\,\,\,\,\,\,\,\,\,\,\,\,\,\,\,\,\,\,\,\,\,\,\,\,\,\,\Big\Vert \sum_{j=1}^N A_{j} A_{j}^{\ast}\Big\Vert_{2\to 2}\Big\}^{\frac{1}{2}}\\
			&\,\,{\leq} \sqrt{\log(n)} \Big(\sum_{j=1}^N \Vert A_{j}\Vert^2_{2\to 2}\Big)^{\frac{1}{2}}
		\end{align*}
		where we used the non-commutative Khintchine inequality due to Lust-Piquard \cite{lust1986inegalites}, \cite{rudelson1999random}. The definition of $B$ and our assumption on $\pi$ in (\ref{main_annahme}) gives
		\begin{align*}
			\sqrt{\log(n)} \Big(\sum_{j=1}^N \Vert A_{j}\Vert^2_{2\to 2}\Big)^{\frac{1}{2}}\leq \sqrt{\log(n)} \,\frac{\sqrt{2\,C_{\Omega,\pi}}}{\sqrt{m}}\,\sqrt{N}.
		\end{align*}
		Now it follows from \cite[Lemma 4.2]{krahmer2014suprema} that 
		\begin{align}\label{eq:cov_numb_maurey_bound}
			\!\!\log \mathcal{N}\Big(\!\text{conv}(B) ,\Vert \!\cdot\!\Vert_{2\to 2}, \!\frac{t}{\sqrt{s}}\Big)\!\lesssim  \! \frac{{s\, C_{\Omega,\pi}}}{m} \!\frac{1}{t^2}{\log (n)}\! \log(4n).
		\end{align}
		Putting (\ref{eq:cov_numb_conv_subset}) and (\ref{eq:cov_numb_maurey_bound}) together yields the first inequality.\par
		To prove the second inequality, first note by inspecting the proof of Lemma \ref{lem:estimate_A_x_2to2} we have
		\begin{align*}
			\Vert A_x-A_y\Vert_{2\to 2}=\Vert A_{x-y}\Vert_{2\to 2}\leq \sqrt{\frac{ C_{\Omega,\pi}}{m}}\Vert x-y\Vert_1.
		\end{align*}
		Therefore, 
		\begin{align*}
			\mathcal{N}(\mathcal{A},\Vert\cdot\Vert_{2\to2},t) &\leq \mathcal{N}\left( D_{s,n},\sqrt{\frac{C_{\Omega,\pi}}{m}} \Vert\cdot\Vert_1,t \right).
		\end{align*}
		Now following the arguments presented in Section 8.4 of \cite{rauhut2010compressive} shows that
		\begin{align*}
			\mathcal{N}\left(\! D_{s,n},\sqrt{\frac{C_{\Omega,\pi}}{m}} \Vert\cdot\Vert_1,t \right)\leq \left(\!1+2\sqrt{\frac{s\, C_{\Omega,\pi}}{m}}\frac{1}{t}\right)^{2s} \!\left(\frac{en}{s}\right)^s.
			\end{align*}
		The second inequality follows immediately.
	\end{IEEEproof}
	
	Now we are able to prove the bound for $\gamma_2(\mathcal{A},\Vert\cdot\Vert_{2\to2})$.
	
	\begin{lemma}\label{lem:abschgamma2}
		We have
		\begin{align*}
			\gamma_2(\mathcal{A},\Vert\cdot\Vert_{2\to2}) \lesssim \sqrt{C_{\Omega,\pi}} \sqrt{\frac{s}{m}} \sqrt{\log (n )\log (4n )}\log(s\, C_{\Omega,\pi}).
		\end{align*}
	\end{lemma}
	\begin{IEEEproof}
		Using the first inequality that was shown in Lemma \ref{lem:covnumbabs} together with Lemma \ref{lem:estimate_A_x_2to2} yields
		\begin{align*}
			&\int\limits_{\frac{1}{\sqrt{m}}}^{d_{2\to2}(\mathcal{A})} \!\!\!\!\sqrt{\log \mathcal{N}(\mathcal{A},\Vert\cdot\Vert_{2\to2},t)} \,\,\text{d}t\\ 
			&\lesssim \int\limits_{\frac{1}{\sqrt{m}}}^{\sqrt{\frac{s\, C_{\Omega,\pi}}{m}}} \!\!\!\!\sqrt{C_{\Omega,\pi}} \sqrt{\frac{s}{m}} \sqrt{\log (n )} \sqrt{\log (4n )} \,\,\frac{1}{t} \,\,\text{d}t\\
			&=\frac{1}{2}\sqrt{C_{\Omega,\pi}} \sqrt{\frac{s}{m}} \sqrt{\log (n )}\sqrt{\log (4n )}\log(s\, C_{\Omega,\pi}).
		\end{align*}
		Performing a similar calculation as presented in Section 8.4 in \cite{rauhut2010compressive} together with the second bound in Lemma \ref{lem:covnumbabs} gives
		\begin{align*}
			&\int\limits_{0}^{\frac{1}{\sqrt{m}}} \sqrt{\log(\mathcal{N}(\mathcal{A}, \Vert \cdot \Vert_{2 \to 2}, t))} \,\,\text{d}t\\
			&\lesssim \sqrt{s} \int\limits_0^{\frac{1}{\sqrt{m}}}\sqrt{ \log\Big(\frac{en}{s}\Big)+\log\Big(1+2\frac{\sqrt{s\, C_{\Omega,\pi}}}{\sqrt{m}\,t}\Big)}  \,\,\text{d}t\\
			&\leq\sqrt{\frac{s}{m}} \bigg( \sqrt{\log\Big(\frac{en}{s}\Big)}+\sqrt{\log\Big(e\big(1+2\sqrt{s\, C_{\Omega,\pi}}\big)\Big)}\,\bigg).
		\end{align*}
		The statement follows from putting both inequalities together. Note that if we were to just use the second bound we would get worse scaling in the exponent of $s$.
	\end{IEEEproof}
		
	With these three bounds at hand, we return to proving Theorem \ref{mainresult}. Let $\delta\in (0,1)$. Using the above lemmata as well as the inequality (\ref{gl:absm}) yields
	\begin{align*}
		E\leq \frac{\delta^2}{c}+\frac{\delta}{\sqrt{c}}+\frac{\delta}{\sqrt{c}}\leq \frac{3\delta}{\sqrt{c}},
	\end{align*}
	where we assumed that $c\geq 1$. Now by choosing the constant $c$ large enough, we get the bound $E\lesssim \frac{\delta}{2c_1}$ where $c_1$ is the absolute constant from Theorem \ref{th:concentration_inequ}. Using that same Theorem then gives
	\begin{align*}
		\mathbb{P}(\delta_s\geq \delta)&\leq \mathbb{P}\Big(\delta_s\geq c_1 E+\frac{\delta}{2}\Big)\\
		&\leq 2 \exp\Big(-c_2 \min\Big\{\frac{0.25\delta^2}{V^2}, \frac{0.5\delta}{U}\Big\}\Big).
	\end{align*}	
	Using the Lemmata \ref{lem:estimate_A_x_2to2}, \ref{lem_estimate_A_x_F} and \ref{lem:abschgamma2} as well as inequality (\ref{gl:absm}) gives the bound
	\begin{align*}
		&2 \exp\Big(-c_2 \min\Big\{\frac{0.25\delta^2}{V^2}, \frac{0.5\delta}{U}\Big\}\Big)\\
		&\leq 2 \exp\Bigg(-c_2 c \frac{0.25}{\big(c^{-\frac{1}{2}}+1\big)^2} \log(\eta^{-1})\Bigg).
	\end{align*}
	Choose the constant $c$ again large enough and deduce that
	\begin{align*}
		2 \exp\Bigg(-c_2 c \frac{0.25 \log(\eta^{-1})}{\big(c^{-\frac{1}{2}}+1\big)^2}\Bigg)\leq 2 \exp\big(-\log(\eta^{-1})\big)
		&=2\eta.
	\end{align*}
	This establishes the theorem.
\end{IEEEproof}

\section{The constant $C_{\Omega,\pi}$}\label{sect:constant}

The dependence of the estimate in Theorem \ref{mainresult} on the choice of representation and sampling set is encapsulated in the constant $C_{\Omega,\pi}$, on which the number of measurements $m$ depends linearly. The determination of this constant (preferably close to 1) remains a challenge. In this section, we present some examples that illustrate this fact.\par
The next proposition establishes that the optimal value $1$ is obtained by the left regular representation $L\colon G\to \text{GL}(\C^G)$, defined by $(L(g)f)(h)=f(g^{-1}h)$.

\begin{proposition}\label{prop:leftregrep_const}
	Let $\Phi$ be constructed by choosing the left regular representation $L$ and as in Definition  \ref{def:measurement_mat}. Then, it holds
	\[\Big\Vert \frac{1}{\sqrt{m}}\, R_{\Omega}\big(L(g) y\big) ^{\ast}_{g\in G} e_h 	\Big\Vert_{2} \leq \sqrt{\frac{1}{m}} \Vert y\Vert_2\]
	for all $y\in\C^G$ and all canonical vectors $e_h\in \C^G$.
\end{proposition}
\begin{IEEEproof}
		For $y\in \C^G$ and a canonical vector $e_h$ we have
		\begin{align*}
			\Big\Vert \frac{1}{\sqrt{m}} R_{\Omega} \big(L(g)y\big)^{\ast}_{g\in G}\, e_h\Big\Vert_2^2 &\leq\frac{1}{m} \sum_{g\in G} \vert \langle e_h,L(g)y\rangle\vert^2\\
			&=\frac{1}{m}\sum_{g\in G} \vert y(g^{-1} h)\vert^2=\frac{1}{m}\Vert y\Vert_2^2.
		\end{align*}
\end{IEEEproof}

It is straightforward to verify that choosing $G=\Z_n$ and $\pi=L$ results in $\Phi$ being a Partial Random Circulant Matrix as examined in \cite{krahmer2014suprema}. So, combining Theorem \ref{mainresult} and Proposition \ref{prop:leftregrep_const} generalizes \cite[Theorem 4.1]{krahmer2014suprema} from $\Z_n$ to arbitrary, even non-abelian, finite groups.

\begin{remark}
	It is possible to generalize Proposition \ref{prop:leftregrep_const} to subrepresentations of the left regular representation $L$ to a certain degree by using more knowledge about group representations than we have presented here.
\end{remark}

To illustrate another advantage of stating Theorem \ref{mainresult} depending on the sampling set as well as the representation, we add one more example: Let $p\in\N$ be prime. We call the set $G=\Z_p\times \Z_p^{\ast}$ together with the operation
\begin{align*}
	(k,l)(k',l'):=(k+lk' \text{ mod } p,ll' \text{ mod } p)
\end{align*}
the affine group. We define a unitary representation of $G$ by
\begin{align*}
	\rho\colon G\to \text{GL}(\C^p),\,(\rho(k,l)f)(j)=f(l^{-1}(j-k))
\end{align*}
for all $(k,l)\in G$. 
\begin{proposition}\label{prop:affine_group}
	Let $\Phi$ be constructed by choosing the representation $\rho$ and as in Definition  \ref{def:measurement_mat}. Then, it holds
	\begin{align*}
		\Big\Vert \frac{1}{\sqrt{m}}\, R_{\Omega}\big(\rho(k,l) y\big) ^{\ast}_{(k,l)\in G} e_j 	\Big\Vert_{2} \leq \sqrt{\frac{\vert\Omega_2\vert }{m}} \Vert y\Vert_2
	\end{align*}
	for all $y\in\C^p$ and all canonical vectors $e_j\in \C^p$ with \linebreak $\Omega_2:=\{l\in\Z_p^{\ast}\mid \exists k\in\Z_p\colon (k,l)\in \Omega\}$.
\end{proposition}
\begin{IEEEproof}
	For $y\in \C^p$ and a canonical vector $e_j$ we have
	\begin{align*}
		&\Big\Vert \frac{1}{\sqrt{m}} R_{\Omega} \big(\rho(k,l)y\big)^{\ast}_{(k,l)\in G}\, e_j\Big\Vert_2^2\leq \frac{1}{m}\sum_{l\in\Omega_2}\sum_{k=1}^p\vert\langle e_j,\rho(k,l)y\rangle\vert^2\\
		& =\frac{1}{m}\sum_{l\in\Omega_2}\sum_{k=1}^p\vert y(lj+k)\vert^2=\frac{\vert\Omega_2\vert }{m} \Vert y\Vert_2^2.
	\end{align*}
\end{IEEEproof}

Since the measurement matrix is known, we can use that knowledge by a priori restricting the sampling set $\Omega$ to be a subset of e.g. $\widetilde{G}=\{(k,1)\in G\,\vert\, k\in\Z_p\}$. Thus, we get $\vert \Omega_2\vert=1$ in Proposition \ref{prop:affine_group} and $C_{\Omega,\rho}=1$.

\begin{remark}
	This idea of a priori restricting the sampling set can actually be generalized to a broader class of representations, namely induced representations of split group extensions. The details will be provided elsewhere.
\end{remark}

We want to finish this section with the following important observation: The availability of sampling sets $\Omega$ with $C_{\Omega,\pi}$ close to 1 can depend on the realization of the representation. For example, the realization of the left regular representation of $\Z_n$ in the Fourier domain gives a measurement matrix $\Phi$ that does not do sparse recovery for any $\Omega$. However, randomizing the sampling set $\Omega$ and slightly adapting $\Phi$ allows for sparse recovery independently of the chosen realization at the cost of slightly larger sampling sets than prescribed in Theorem \ref{mainresult}.\par
Understanding this phenomenon, as well as further questions that are linked to the analysis of the constant $C_{\Omega,\pi}$, is still ongoing work and will be presented elsewhere.

\section*{Acknowledgment}
We thank H. Rauhut (LMU Munich) for fruitful discussion and advice.
\bibliographystyle{IEEEtran}
\bibliography{IEEEabrv,IEEEreferences25}

\end{document}